%% file: main.tex
\documentclass[sigconf, screen=true, nonacm]{acmart}

\usepackage{algorithm}
\usepackage{bbm}
\usepackage{caption,subcaption}
\usepackage{algpseudocode}

\DeclareMathOperator*{\argmax}{arg\,max}

\AtBeginDocument{%
  \providecommand\BibTeX{{%
    \normalfont B\kern-0.5em{\scshape i\kern-0.25em b}\kern-0.8em\TeX}}}

\setcopyright{acmcopyright}
\copyrightyear{2021}
\acmYear{2021}
\acmMonth{4}

\acmConference[WWW'21]{WWW'21: The Web Conference 2021}{April 19--23, 2021}{Ljubljana, Slovenia}
\acmBooktitle{Proceedings of The Web Conference 2021 (WWW'21),
April 19--23, 2021, Ljubljana, Slovenia}

\acmSubmissionID{1244}

\begin{document}

\title{Typing Errors in Factual Knowledge Graphs: Severity and Possible Ways Out}
\titlenote{Camera-ready for 30th The Web Conference (WWW'2021)}

\author{Peiran Yao}
\orcid{0000-0001-5300-1734}
\affiliation{%
  \institution{University of Alberta}
  \streetaddress{2-21 Athabasca Hall}
  \city{Edmonton}
  \state{AB}
  \country{Canada}
  \postcode{T6G 2E8}
}
\email{peiran@ualberta.ca}

\author{Denilson Barbosa}
\orcid{0000-0001-8017-2096}
\affiliation{%
  \institution{University of Alberta}
  \streetaddress{2-21 Athabasca Hall}
  \city{Edmonton}
  \state{AB}
  \country{Canada}
  \postcode{T6G 2E8}
}
\email{denilson@ualberta.ca}


\begin{abstract}
  Factual knowledge graphs (KGs) such as DBpedia and Wikidata have served as part of various downstream tasks and are also widely adopted by artificial intelligence research communities as benchmark datasets. However, we found these KGs to be surprisingly noisy. In this study, we question the quality of these KGs, where the typing error rate is estimated to be 27\% for coarse-grained types on average, and even 73\% for certain fine-grained types. In pursuit of solutions, we propose an active typing error detection algorithm that maximizes the utilization of both gold and noisy labels. We also comprehensively discuss and compare unsupervised, semi-supervised, and supervised paradigms to deal with typing errors in factual KGs. The outcomes of this study provide guidelines for researchers to use noisy factual KGs. To help practitioners deploy the techniques and conduct further research, we published our code and data \footnote{\url{https://github.com/xavieryao/kg-type-err-corr}}.
\end{abstract}



\maketitle
\renewcommand{\shortauthors}{Yao and Barbosa}

\input{introduction.tex}

\input{severity.tex}

\input{spectrum.tex}

\input{experiment.tex}
\input{review.tex}
\input{conclusion.tex}

\begin{acks}
The authors would like to thank J. Lau for the discussions on problem formulation. We would also like to thank the anonymous reviewers for their generous and valuable feedback. This work is supported by the \grantsponsor{nserc}{Natural Sciences and Engineering Research Council of Canada}{https://www.nserc-crsng.gc.ca} and the \grantsponsor{scotiabank}{Scotiabank Artificial Intelligence Research Initiative}{https://www.scotiabank.com/}.
\end{acks}

\bibliographystyle{ACM-Reference-Format}
\bibliography{references}


\end{document}

%% file: introduction.tex
\section{Introduction}
Large scale factual knowledge graphs (KGs) such as DBpedia \cite{lehmann2015dbpedia} and Wikidata \cite{vrandevcic2014wikidata} organize factual knowledge extracted from trustworthy corpus like Wikipedia in a machine-readable way. As an accessible and effective source of information, they have served as a vital component of many AI systems including question-answering systems \cite{lukovnikov2017neural,huang2019knowledge}, recommendation systems \cite{cao2019unifying}, and contextualized language models \cite{peters2019knowledge}. Aside from being utilized by a variety of downstream applications, in recent years these factual KGs have also been widely adopted as a benchmark in a multitude of research on knowledge graphs or even general machine learning. DBpedia and Wikidata have been used to evaluate KG embeddings \cite{ristoski2016rdf2vec}, few-shot link prediction \cite{xiong2018one}, entity typing \cite{xu2018neural}, and other tasks. Outside of the research area of knowledge graphs, these KGs are also used as datasets for general ML tasks such as semi-supervised text classification \cite{dai2015semi, miyato2016adversarial,xie2019unsupervised}.

Although being profoundly favored by the AI research community, the quality of these factual KGs, however, might be questionable. In fact, our analysis shows that even on the coarse level, the percentage of typing errors in DBpedia has already reached 27\%. This number suggests that DBpedia, as well as other KGs similarly built, are quite noisy. The facts in DBpedia were extracted automatically from Wikipedia based on a collectively-maintained set of rules, so it is not surprising that the extracted facts contain errors. This also indicates that more caution is needed when using factual KGs, and it of great importance to develop methods to identify typing errors in KGs.

In this study, we propose to use semi-supervised noise model to effectively detect typing errors with only a minimum amount of human intervention. We designed a neural network (NN) architecture specially for entity typing in factual KGs, which combines heterogeneous information from entity descriptions, surface forms of entity names, and network structures of the KG. In addition to that, we included a probabilistic noise model to enable the model to robustly learn from entities with noisy type labels, and used virtual adversarial training \cite{miyato2016distributional} to learn from all entities disregarding whether the label is noisy or not. We also applied an active learning strategy to only annotate the most useful entities.

Data-driven approaches to deal with typing errors in factual KGs have a very broad spectrum, covering fully unsupervised clustering and outlier detection \cite{oza2018one, aly2019every}, semi-supervised noise models that could leverage noisy labels \cite{kremer2018robust, goldberger2016training, jindal2019effective}, and supervised noise detection methods that fully rely on gold labels \cite{caminhas2019detecting, zhou2017dbpedia}. In this study, we present a taxonomy of the KG typing error detection paradigms and comprehensively evaluate those paradigms on DBpedia.

To the best of our knowledge, this work is the first to apply noise models and semi-supervised learning to resolve typing errors in factual KGs. Although the theory of noise models have been developed for years, they were mostly evaluated \emph{in vitro} on synthesized datasets and this is also one of the earliest attempts to test them on a real noisy dataset. The findings of our study reveal the practical difficulties when applying those models in reality and provide directions on dealing with KG typing errors. Despite the extensive effort we made to develop error detection methods with multiple paradigms, the problem remains largely unsolved. The code and data used in the study will be released to the public after the publication of this paper to encourage researchers to deploy the techniques by their needs and further the study.

%% file: severity.tex
\begin{figure}
    \centering
    \includegraphics[width=0.46\textwidth]{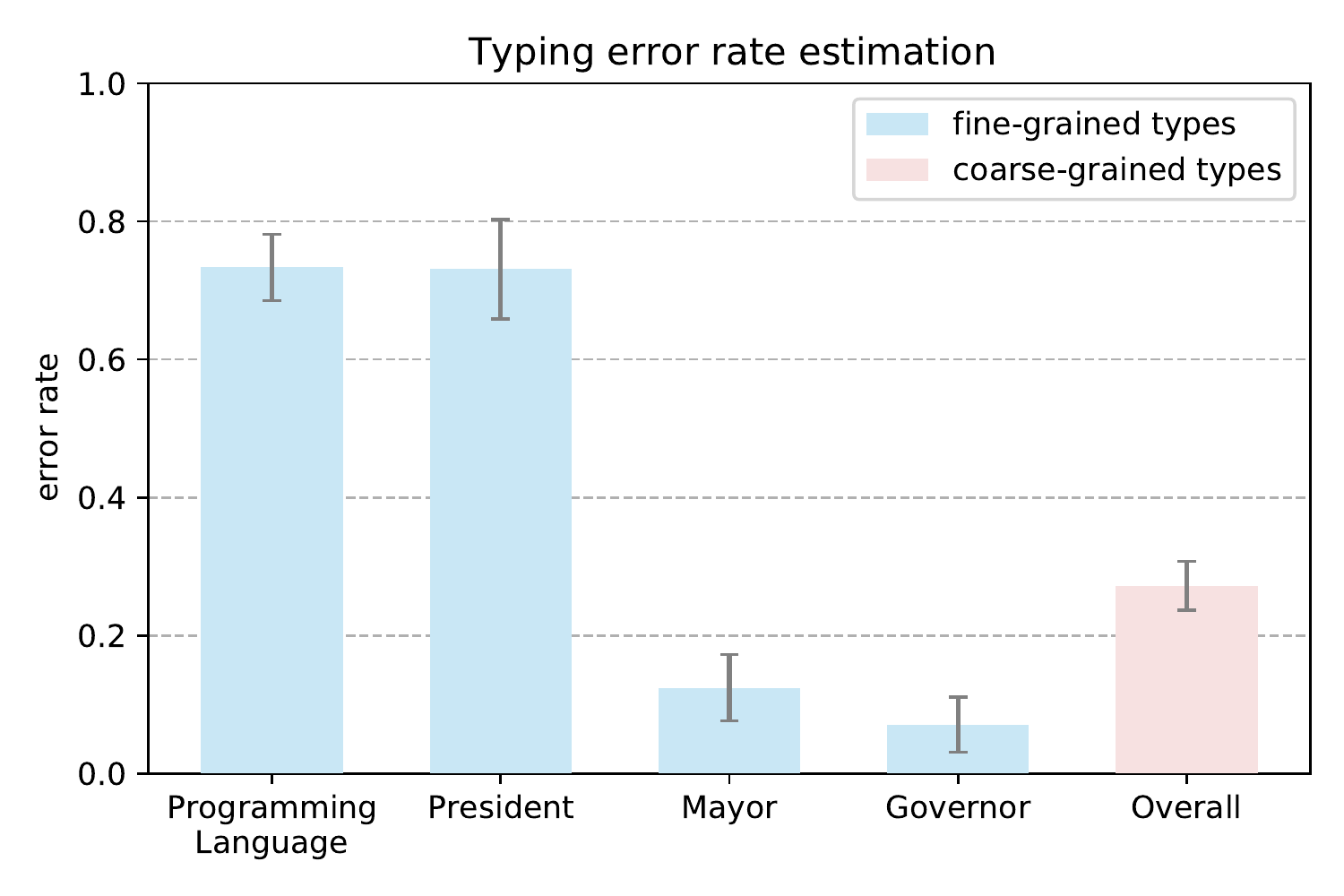}

    \caption{Estimations of fine-grained and coarse-grained error rates in DBpedia. The left 4 bars (in blue) show the error rates of 4 fine-grained types, and the right-most bar (in red) shows the mean error rate of coarse-grained entity types.}
    \Description{ProgrammingLanguage: $0.733\pm 0.048$, President: $0.731 \pm 0.072$, Mayor: $0.124 \pm 0.048$, Governor: $0.071\pm0.04$, overall: $0.272 \pm 0.0355$}

    \label{fig:error-ratio}
\end{figure}

\section{Typing Errors in Knowledge Graphs}
\subsection{Problem Definition}
Factual KGs like DBpedia are sets of triples like \texttt{<head, relation, tail>}, where \texttt{head} and \texttt{tail} are entities. For typing errors we are interested in the \texttt{rdf:type} relation, for example the following tuples regarding the type of \texttt{dbr:Canada} are present in DBpedia: 
\begin{verbatim}
    <dbr:Canada, rdf:type, dbo:Place>
    <dbr:Canada, rdf:type, dbo:PopulatedPlace>
    <dbr:Canada, rdf:type, dbo:Country>
    <dbr:Canada, rdf:type, dbo:MusicalArtist>
\end{verbatim}
where \texttt{<dbr:Canada, rdf:type, dbo:MusicalArtist>} is a typing error.

Entity types are also organized hierarchically as a rooted tree in the DBpedia Ontology. We use $level(y)$ to denote the level of the type $y$ in the rooted tree, and the level of the root type, \texttt{dbo:Thing}, is $0$. A coarse-grained type $y_{c}$ satisfies $level(y_{c}) = 1$. We denote the correct type of an entity $e$ as $\hat{y}$. 

We define the problem of \textbf{KG typing error detection} as: given a tuple \texttt{<e, rdf:type, y>} from a KG, determine whether it is correct or incorrect. Moreover, if we restrict $level(y) = 1$, the problem becomes \textbf{coarse-grained KG typing error detection}.

\subsection{Error Rates in DBpedia}
\label{sec:error-rate}
Zaveri et al. estimated that about 12\% of the tuples in DBpedia are erroneous \cite{zaveri2013user}. When it comes to entity types, the issue seems worse. We estimated the ratio of fine-grained typing errors in DBpedia by looking at four fine-grained types, and the ratio of coarse-grained typing errors by examining a subset of entities. The estimated ratios and the confidence intervals with 95\% confidence are shown in Figure \ref{fig:error-ratio}. 

The results shown that even with the coarsest granularity, the overall error rate is already $0.272 \pm 0.0355$. This is the lower bound of errors, and with finer granularity the situation should be equal or even worse. Another finding is, there are both types of good quality and poor quality. For example, the types \texttt{Mayor} and \texttt{Governor} have error rates less than $0.15$ while the rate is as high as $0.73$ for \texttt{ProgrammingLanguage}. This makes it challenging to develop a universal method for error detection and correction. These statistics were estimated on DBpedia version 2016-10. We chose this version as this version is widely used by the research communities and has a relatively large amount of public resources such as analyses, baseline methods, and pre-trained models available as a result.

\subsection{Issues with the DBpedia Taxonomy}
\label{sec:dbpedia-taxonomy-issues}
Despite the problems with the data acquisition process such as imperfect rules, we also identified issues with the DBpedia Taxonomy design which might have induced typing errors. For the type hierarchy, certain types are not positioned correctly, and there are also overlaps between types. Certain types also have ambiguous names and lack formal definitions.

%% file: spectrum.tex
\begin{figure*}[bt]
    \centering
    \includegraphics[width=0.75\linewidth]{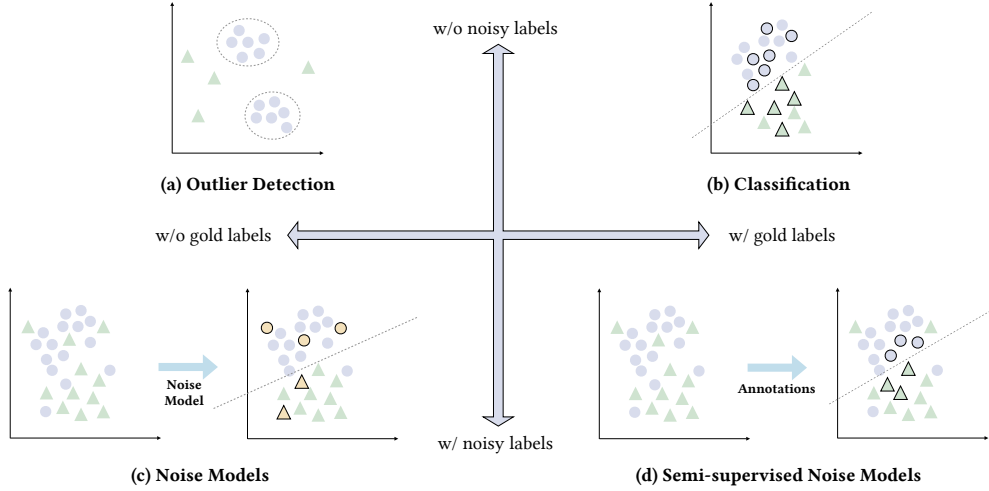}
    \caption{The spectrum of typing error detection paradigms. Triangles (green) and circles (purple) denote two distinct types. In (b) and (d), shapes with solid borders represent entities with gold type labels. In (c), shapes in yellow and with solid borders represent entities whose label was altered by the noise model.}
    \label{fig:spectrum}
    \Description{The illustration of the specturm of typing error detection paradigms. (a) Top-left (outlier detection): without noisy labels and without gold labels. Normal instances (more concentrated) are separated from outliers (at a distance from major clusters). (b) Top-right (classification): without noisy labels but with gold labels. A seperation plane is obtained from gold labels to separate correct and wrong instances. (c) Bottom-left (noise models): without gold labels but with noisy labels. Some labels are flipped by the noise model, and a separation plane is obtained after the noise model manipulates the labels. (d) Bottom-right (semi-supervised noise models): with gold labels and with noisy labels. Some labels are flipped by huamn annotators, some are flipped by the noise model. A separation plane is obtained after these manipulations.} 
\end{figure*}

\section{The Spectrum}

We categorized the possible paradigms for typing error detection based on the amount of intrinsic and extra information required. As illustrated in Figure \ref{fig:spectrum}, the approaches are positioned in four quadrants, in terms of whether they utilized the noisy type labels and whether they require additional annotations. In this section we will briefly introduce the ideas behind each paradigm and propose our methods for KG typing error detection adopting the paradigms. 

\subsection{Noise Model for Error Detection}
Inspired by the recent advances in learning with noise \cite{kremer2018robust, goldberger2016training, jindal2019effective}, we propose to use the combination of an entity typing network and a probabilistic noise model for the typing error detection task. We trained a classification model that is robust to noise with a subset of a noisy KG, and applied that model on another subset to detect errors. Learning with noise enables us to leverage the vast amount of data in noisy KGs, without the need of human labour to obtain high-quality typing labels.

\subsubsection{Entity Typing Network}

\begin{figure*}[t]
    \centering
    \includegraphics[width=0.7\linewidth]{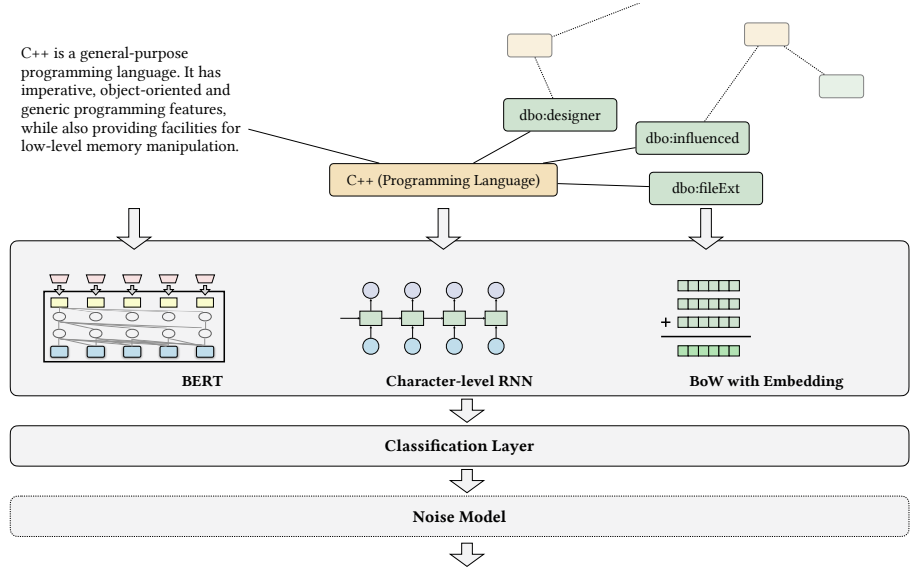}
    \caption{The architecture of the entity typing network. We use the entity \texttt{C++ (ProgrammingLanguage)} in DBpedia as the example here. Green nodes (starting with \texttt{dbo:}) represent relation nodes in the KG, and orange nodes (not starting with \texttt{dbo:}) represent entities in the KG. Example texts are from DBpedia  (http://dbpedia.org/page/C++).}
    \label{fig:architecture}
    \Description{Top layer (input). A combination of text (for example, "C++ is a general-purpose programming language. It has imperative, object-oriented and generic programming features, while also providing facilities for low-level memory manipulation."), surface form (for example, "C++ (Programming Language)"), and relation nodes (for example, "dbo:designer", "dbo:influenced", and "dbo:fileExt"). Second layer: BERT, character-level RNN and bag-of-word with embedding. Third layer: a classification layer. Forth layer: the noise model. More layers to follow.}
\end{figure*}

To leverage the heterogenous information present in factual KGs, we designed a neural network architecture for entity representation learning, as shown in Figure \ref{fig:architecture}. The description of an entity is a rich source of typing information, as for entity typing simple pattern matching could achieve 87\% accuracy \cite{kliegr2015linked}. The network captures this information with a pre-trained BERT model \cite{devlin2019bert}. The name or surface form of an entity  could often suggest its type, for example entities whose name ends with "\emph{Script}" are more likely to be programming languages. Therefore, surface forms are encoded with a character-level RNN in our model. And finally, the network captures the first-order network structure of the KG with a bag-of-word (BoW) model. Each relation in the KG is represented as an embedding vector $\vec{r}$, and the representation of the graph structure is the sum of the embeddings of all relations an entity has. Formally, for an entity $e$ in a KG $G$, the typing network works as follows, where $\vec{r}$, $\mathbf{W}$ and $\vec{b}$ are parameters to be learned:

\begin{equation}
    \vec{e} = \left[ BERT(description); RNN(name); \sum_{r: (e, r, e' \in G)} \vec{r}\right]
\end{equation}
\begin{equation}
    \vec{o} = ReLU(\mathbf{W}^T\vec{e} + \vec{b})
\end{equation}

Each component of the network (BERT for description, character-level RNN for surface form, and BoW for graph structure) was initialized independently by pre-training with an entity classification task on a noisy entity type dataset.

\subsubsection{Noise Model}
Since our goal is to detect errors, we choose to train a multi-label classification model instead of a multi-class one, to reduce complexity. The probability of the entity $e$ having type $z_i$ given by the typing network is as follows where $\Theta$ is the parameters of the model.
\begin{equation}
    \Pr(z_i|e; \Theta) = \frac{1}{1 + e^{-\vec{o}_i}}
\end{equation}
Following recent developments in noise models \cite{kremer2018robust, goldberger2016training, jindal2019effective}, we use the probabilities of flipping to model the process of typing error generation:
\begin{equation}
    p_i \triangleq \Pr(y_i|z_i)
\end{equation}
where $p_i$, $i = 1 \cdots T$ are the parameters of the noise model that are learned end-to-end during the training of the model, and $T$ is the total number of types.

The final output after the noise model is then
\begin{equation}
    \Pr(y_i | e; \Theta) = p_i \Pr(z_i|e; \Theta) + (1-p_i)\left(1-\Pr(z_i|e; \Theta)\right)
\end{equation}

\subsection{Semi-supervised Noise Model}
Our model was further extended to learn simultaneously from entities with noisy typing labels and those with human-verified gold labels. We used a two-fold approach here. First, we applied virtual adversarial training (VAT) \cite{miyato2016distributional} as a regularization method, which makes use of the input data disregarding the labels, and hence avoids being affected by the typing errors in the labels. When calculating the loss, we used the model prediction without noise model $\Pr(z_i|e)$ instead of $\Pr(y_i|e)$ if an entity has gold label, and we proposed an active learning scheme to efficiently select entities to annotate. The learning rate is also dynamically adjusted based on the prior belief of the correctness of a type label estimated from word embeddings.

\subsubsection{Virtual adversarial training}

We applied VAT to learn from all sample entities in the dataset without being affected by erroneous type labels. VAT ensures a soft constraint that the model predictions for similar input entities are the same. The similarity of entities is measured in the embedding space by the $L_2$ distance of embedding $\vec{e}$. For an input entity $e$ and its embedding $\vec{e}$, VAT adds the following local distributional smoothing (LDS) term to the loss function:
\begin{equation}
    LDS(\vec{e}) \triangleq - \Delta_{KL}(\vec{r}_{e}, \vec{e})
\end{equation}
where 
\begin{align}
    \Delta_{KL}(\vec{r}, \vec{e}) &\triangleq KL[\Pr(y|\vec{e})||\Pr(y| \vec{e} + \vec{r}_e)] \\
    \vec{r}_e &\triangleq \argmax_{r: ||r||_2 \leq \epsilon}  \Delta_{KL}(\vec{r}, \vec{e}) 
\end{align}
and $\epsilon$ is a hyper-parameter.

We denote the set of entities with only noisy type labels as $S = \{(e^{(i)}, y^{(i)})\}_{i:1\cdots N}$ and the set of entities with gold labels as $\hat{S} = \{(e^{(i)}, \hat{y}^{(i)})\}_{i:1\cdots M}$. The adjusted loss function that considers both noisy label and label as well as VAT is:
\begin{equation}
\begin{split}
    \label{eq:loss}
    L(S, \hat{S}; \Theta) &= \\
    \mathop{\mathbb{E}}_{(e, y) \in S \cup \hat{S}} \Big[  &\mathbbm{1}[(e, y) \in S] \cdot \sum_{i=1}^{T}BCE(\Pr(y_i|e; \Theta), \mathbbm{1}(y=y_i)) \\
    &+\mathbbm{1}[(e, y) \not\in S] \cdot \sum_{i=1}^{T}BCE(\Pr(z_i|e; \Theta), \mathbbm{1}(y=y_i)) \\
    &+  \lambda \cdot LDS(\vec{e})  \Big]
\end{split}
\end{equation}
where $\lambda$ is the hyper-parameter controlling the impact of VAT and $\Theta$ is the parameters of the typing network and the noise model.

\subsubsection{Active entity selection}
We propose to use uncertainty sampling (\textbf{US}) as the active learning strategy for selecting entity and type pairs to annotate. Although it is not theoretically optimal, it requires less computation and is thus more scalable on large KGs. Uncertainty sampling selects the entity whose prediction has the maximum entropy:
\begin{equation}
    \label{eq:us}
    e_{sel}, y_{sel} = \argmax_{(e, y)} \sum_{i=1}^T H_b(\Pr(y_i|e; \Theta))
\end{equation}

We also compared US with expected error reduction (\textbf{ERR}) \cite{roy01towardoptimal,kremer2018robust}. In ERR, samples are selected greedily to maximize the expected model change, and this process is approximated by the difference between the gradients before and after an annotation. Suppose $(e_{sel}, y_{sel})$ is selected for annotation and the obtained gold label is $\hat{y_{sel}}$, the loss function after the annotation is then
\begin{equation}
    \tilde{L}(e_{sel}, y_{sel}; \Theta) \triangleq  L\left(S \backslash \{(e_{sel}, y_{sel})\}, \hat{S}\cup\{  (e_{sel}, \hat{y}_{sel}) \}; \Theta \right)
\end{equation}

The entity-label pair selected for annotation is then the one that maximizes the gradient difference:
\begin{equation}
    \label{eq:err}
    e_{sel}, y_{sel} = \argmax_{(e, y)} \left\lVert \frac{\partial L}{\partial\Theta} - \frac{\partial \tilde{L}(e, y)}{\partial\Theta}  \right\rVert
\end{equation}

\subsubsection{Dynamic learning rate}
We adjusted the learning rate for each entity in the training set according to the prior belief of the probability that the entity has correct type labels. This encourages the model to learn more from entities with correct labels and mitigates the negative impacts of noisy labels. For an entity-type pair $(e, y)$, we estimate the prior probability that the label $y$ is correct to be the cosine similarity between the GloVe \cite{pennington2014glove} embeddings of the names of the entity and the type, denoted as $\vec{w}_e$ and $\vec{w}_y$. If one of the two word embeddings does not exist, the prior probability falls back the the mean probability of all entities. Suppose the original learning rate is $lr$, then the dynamic learning rate for the entity-type pair $(e, t)$ is set to be within the range of $[0.5lr, 1.5lr]$ with the formula:
\begin{equation}
    lr_{dyn}(e, y) = \left( 0.5 + \frac{1}{2}\cdot\cos(\vec{w}_e, \vec{w}_y) \right)\cdot lr
\end{equation}

The complete framework of the semi-supervised typing error detection method we proposed is described in Algorithm \ref{alg:ss-err-detection}.

\begin{algorithm}[H]
    \caption{Training the semi-supervised error detection model}
    \label{alg:ss-err-detection}
    \begin{algorithmic}[lines]
        \State $\hat{S} \gets \emptyset$ \Comment{$\hat{S}$ is the gold training set}
        \Procedure{Epoch}{$S$, $\hat{S}$} \Comment{$S$ is the noisy training set}
            \For{$batch \in S \cup \hat{S}$}
                \State optimize $L(S, \hat{S}; \Theta)$ with $batch$ \Comment{$L$ defiend in (\ref{eq:loss})}
                \For {$i \in 1 \cdots MaxQuery$}
                    \State select $(e_{sel}, y_{sel}) \in S$ based on (\ref{eq:err})
                    \State annotate $e_{sel}$ with gold label $\hat{y}$
                    \State $S \gets S \backslash \{(e_{sel}, y_{sel})\}$; $\hat{S} \gets \hat{S} \cup \{(e_{sel}, \hat{y})\}$
                \EndFor
            \EndFor
        \EndProcedure
    \end{algorithmic}
\end{algorithm}

\subsection{Outlier Detection}
Outlier detection methods are the least costly to deploy for typing error detection, as they require no label. For typing error detection, outlier detection algorithms are independently applied on individual types, and the input is the embeddings $\vec{e}$ of entities labelled with type $t$ in the KG. The assumption behind outlier detection algorithms is entities with correct types could form high density clusters \cite{scikit-learn}. But this assumption is often violated because entities with correct labels are not the majority for certain types, as discussed in Section \ref{sec:error-rate}.

\begin{figure}[ht]
    \centering
    \includegraphics[width=0.95\linewidth]{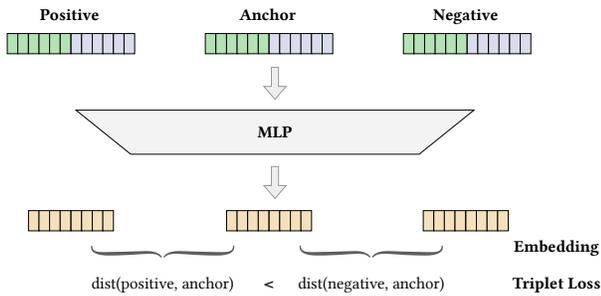}
    \caption{The representation learning process used to reduce the dimensionality of the embedding for outlier detection.}
    \label{fig:repr-learning}
    \Description{
        The input vectors of positive, anchor, and negative entities are fed into an MLP. The MLP produces learned embeddings for the three entities. The triplet loss enforces that the distance between the positive and anchor embeddings is smaller than the distance between the negative and anchor embeddings.
    }
\end{figure}

We used the combination of Wikipedia2Vec \cite{yamada2020wikipedia2vec} and RDF2Vec \cite{ristoski2016rdf2vec} embeddings for each entity as the input. These two embedding methods do not require labels, either manually-provided or noisy, as they are self-supervised. The Wikipedia2Vec captures text information from the entity descriptions while RDF2Vec captures information from the graph structure of the KG. And to reduce the dimensionality of embeddings, we trained a MLP-based representation learning network with triplet loss, as illustrated in Figure \ref{fig:repr-learning}. The input of the MLP is the concatenation of the Wikipedia2Vec and RDF2Vec embedding of an entity, and the output is the embedding with reduced dimensionality. For each entity-type pair, an anchor entity is sampled from the set of positive samples for the type and a negative entity is sampled from negative samples. The triplet loss imposes a constraint that the positive entity should be closer to the anchor than the negative one, measured by the cosine distance of the embeddings. The outlier detection algorithms we used are Local Outlier Factor (\textbf{LOF}) \cite{breunig2000lof} and Isolation Forest (\textbf{IF}) \cite{liu2008isolation}.

\input{coarse-results.tex}

\subsection{Classification}
\label{sec:classification}
This paradigm requires to provide gold type labels for entities and train binary or multi-class classifiers to classify entities to the right types. Some previous studies \cite{caminhas2019detecting,zhou2017dbpedia} proposed to use supervised classification for typing error detection, but it is hard to scale as the number of types is large for many KGs (in total 778 types in DBpedia). One work \cite{caminhas2019detecting} tried to tackle the scalability problem by another entity type dataset of better quality, but this could not fundamentally solve the issue as external datasets are also noisy and may be unavailable. Due to these concerns, we omit this paradigm.

%% file: coarse-results.tex
\begin{table*}[t]
    \caption{Results of coarse-grained typing error detection.}
    \label{tab:coarse-results}
    \begin{tabular}{l||ccc|ccc|ccc}
      \toprule
      Model & \multicolumn{3}{c|}{80 annotations / epoch 1} & \multicolumn{3}{c|}{140 annotations / epoch 2} & \multicolumn{3}{c}{200 annotations / epoch 3}\\
      & Prec. & Recall & F1 & Prec. & Recall & F1 & Prec. & Recall & F1 \\
      \midrule
      \textbf{SSNM (US)}          & 0.83 & 0.63 & \textbf{0.72} & 0.82 & 0.63 & \textbf{0.71} & 0.74 & 0.55 & 0.63\\
      \textit{- VAT}              & 0.84 & 0.62 & 0.71 & 0.85 & 0.59 & 0.69 & 0.78 & 0.53 & 0.63\\
      \textit{- VAT, -dynamic lr} & 0.76 & 0.57 & 0.65 & 0.77 & 0.52 & 0.62 & 0.75 & 0.47 & 0.58\\
      \textit{- gold label}       & 0.64 & 0.47 & 0.54 & 0.67 & 0.43 & 0.52 & 0.67 & 0.43 & 0.52\\
      \textit{- gold label, - NM} & 0.67 & 0.45 & 0.54 & 0.64 & 0.43 & 0.51 & 0.65 & 0.43 & 0.52\\
      \textit{- noisy data, -US}  & 0.27 & 0.50 & 0.35 &  -   &  -   &  -   &  -   &  -   & -   \\
      \textit{- noisy label, -US} & 0.44 & 0.82 & 0.57 & 0.50 & 0.80 & 0.62 & 0.69 & 0.72 & \textbf{0.70}\\
      SSNM (ERR)                  &      &      &      &      &      &      &      &      &     \\
      \textit{- VAT}              & 0.77 & 0.52 & 0.62 & 0.73 & 0.52 & 0.61 & 0.72 & 0.46 & 0.56\\
    \bottomrule
  \end{tabular}
\end{table*}

%% file: experiment.tex
\section{Experiments}

\subsection{Dataset}
We chose DBpedia \cite{lehmann2015dbpedia} as the factual KG of interest, because DBpedia is still actively developed and has been widely used in the research community. The datasets used for our experiments were derived from the DBpedia version 2016-10 because of the popularity of this version and the large amount of resources. We prepared a large-scale coarse-grained typing dataset to test our proposed methods based on noise models, and adapted the dataset from the thesis of Caminhas \cite{caminhas2019detecting} as a fine-grained typing dataset to test outlier detection and classification methods.

\subsubsection{Coarse-grained typing dataset}
This dataset (\textbf{DBpedia-C}) is a subset of DBpedia with balanced type distribution, and has a coarse typing granularity. This dataset was created because our methods based on noise models follow a multi-label classification setting, and keeping only the coarse-grained types could reduce the complexity of the task. For each entity we kept its most general type in the DBpedia type ontology, which consists of 17 distinct types in total, including one (\texttt{Other}) for all minority types. There are 56 types in the first (most general) level of the DBpedia type hierarchy, and for the types with more than 10,000 entities we uniformly sampled a subset with size 10,000, and those types with less than 10,000 entities were aggregated as the \texttt{Other} type and that type was also subsampled to a subset with size 10,000. The final dataset was uniformly subsampled to a size of 500,000. We followed a 97:3 train-dev split, where the dev set was used to tune the hyper-parameters of the model. In addition, we have a test set with 600 entities with manually annotated gold labels for evaluation. And during the active learning process, additional 3,247 entities were annotated with gold labels.

\subsubsection{Fine-grained typing dataset}
We adapted the dataset from the work by Caminhas \cite{caminhas2019detecting} (denoted as \textbf{DBpedia-F}) to evaluate outlier detection and classification methods, as they are conducted on a type-by-type basis. The \textbf{DBpedia-F} dataset contains 83 fine-grained types from DBpedia with 5,889 positive samples and 3,395 negative samples. The positive and negative samples were obtained by sampling and examining a subset of entities for each type. Note that the sampling process was not uniform, so this dataset could not represent the true noise ratio in DBpedia. The dataset was divided by types to create the train/dev/test split, and each split contains 48, 16, and 19 types respectively.

\subsection{Implementation Details}
We implemented the entity typing network, the semi-supervised noise model and the representation learning model for outlier detection with PyTorch. In the typing network, the BERT component we used is the \texttt{bert-base-uncased} model from HuggingFace\footnote{\url{https://huggingface.co/transformers/pretrained_models.html}}. We preprocessed the input entity descriptions with the \texttt{en\_core\_web\_sm} NER model from spaCy\footnote{\url{https://spacy.io}} by replacing all named entities with a special token \texttt{ENT} and all locations with \texttt{LOC}. The character-level RNN for surface forms is a uni-directional RNN with hidden size 64. For the BoW model for graph structures, we kept all nodes with more than 20 occurrences and used a hidden size of 256. When training the typing network, we used a hidden size of 512, a batch size of 128, and set $\lambda$ to be $0.1$ after tuning on the development set. The parameters $p_i$ for the noise model were all initialized to be $1.0$. We used the Adam optimizer with learning rate \emph{1e-3}, and used the pre-trained \texttt{6B.100d} GloVe embedding for dynamic learning rate \cite{pennington2014glove}. We applied the active learning strategy to label a batch of 20 entities for every 400 iterations, which is around 70 annotations per epoch.

The outlier detection algorithms from \emph{scikit-learn} were used. We used the pre-trained \texttt{enwiki\_20180420} Wikipedia2Vec embeddings with 100 dimensions \cite{yamada2020wikipedia2vec} and the \texttt{uniform} RDF2Vec embeddings with 200 dimensions \cite{ristoski2016rdf2vec}, and reduced the dimensions with our representation learning network to 128 dimensions.

\subsection{Evaluation of Noise Models}
\subsubsection{Results}
We evaluated the proposed semi-supervised noise model (\textbf{SSNM}) on the task of coarse-grained typing error detection with the \textbf{DBpedia-C} dataset. The performance of SSNM as well as several baselines are shown in Table \ref{tab:coarse-results}. The results at different stages of the active training process are reported to compare the effect of training iterations. For models not involving active learning (\emph{SSNM(US) -noisy label, -US} and \emph{SSNM(US) -noisy data, -US}), we only limited the number of annotated entities and reported the results of the checkpoints with best validation accuracy in the first 10 epochs.

The results show that our model achieved very high F1 score ($0.72$) with only 80 gold labels, and the F1 scores are above all other baselines when there are only 80 and 140 gold labels. This is a strong indication of the efficiency and effectiveness of our model.

\subsubsection{Effects of information sources}
We assume that our proposed method achieved good performance by leveraging information from various information sources, including the entity data points, gold labels, and noisy labels. We verified this assumption by comparing its performance with a few ablated baselines. The \emph{-noisy data, -US} baseline only used the limited number of gold-labelled entities to train a fully-supervised classifier, and it achieved the poorest F1 score ($0.35$ v.s. $0.72$). This indicates that only relying on gold labels is infeasible if the gold-labelled set is small. The \emph{-gold label} and \emph{-gold label, -NM} baselines only rely on entities with noisy typing labels, and the performance is not very satisfying ($0.54$ v.s. $0.72$). This justifies the need for additional human annotations. The \emph{-noisy label, -US} baseline uses entities with and without gold labels, but treat entities without gold labels as unlabelled. Compared with the \emph{-noisy data, -US} baseline it has a huge performance gain ($0.57$ v.s. $0.35$), which also suggests the usefulness of noisy data. Incorporating prior belief of label correctness from self-supervised embeddings with our dynamic learning rate scheme (\emph{-VAT} v.s. \emph{-VAT, -dynamic lr}) also has positive impacts. Finally, the \emph{-VAT} baseline uses less information from the noisy data points and suffered from a performance drop, which suggests that entities with noisy labels are helpful if we ignore the labels.

\subsubsection{Comparison of active learning strategies}
We compared the performance of uncertainty sampling (US) and error rate reduction (ERR) as the active learning strategy for sampling entities to annotate. The evaluation shows that uncertainty sampling, though theoretically imperfect, achieved better results than ERR. And in practice, ERR is computationally intensive as it involves computing the gradient as many times as the number of input entities. This makes it infeasible to iterate through every entity in the dataset, and makes the sampling less accurate. When used in noise models, ERR requires two distinct loss functions for gold label and noisy label. In our case, the difference between the two loss functions is whether the noise model is applied or not, which might not work well as in the early stage of training the noise model has not started to take effects.

\subsubsection{Effects of training iterations}
When the number of training iterations is large (at epoch 3), although more entities are annotated with gold labels, the performance of all models involving noisy labels began to decay (for the \emph{SSNM(US)} model, F1 drops from $0.72$ at epoch 1 to $0.71$ at epoch 2 and $0.63$ at epoch 3). We suspect that this is because as the training proceeds, the model begins to fit more on the wrong labels. This coincides with the previous finding of Arazo el al. that samples with wrong labels are often hard to learn and are learned at a later stage of the training process \cite{arazo2019unsupervised}. Although we applied techniques such as fine-tuning on the entities with gold labels, this issue was not completely mitigated. In the meantime, the \emph{-noisy label, -US} baseline does not have this issue as it ignores noisy labels, and therefore achieved the best performance at epoch 3.

\subsubsection{Ablation study of the entity typing network}
The entity typing network we designed combines heterogeneous information from entity description, entity name (surface form), and the KG structure (neighboring nodes of the entity in the KG). We assessed the contributions of each of these three features with an ablation study, where we used the typing network to perform entity type classification on the DBpedia-C dataset without gold labels. This coincides with the setting of typical entity typing or classification tasks where noisy typing labels are ignored. We report the best accuracy and loss on the dev set of the first 10 epochs, and the test accuracy at this best epoch. Ablation was achieved by replacing the corresponding part of the input embedding to random noise generated from a uniform distribution in range $[0.0, 1.0]$ and keeping the others parts unmodified.

\begin{table}[htb]
  \caption{Classification accuracy and loss of the entity typing network on the dev set and the test set.}
  \label{tab:nn-results}
  \begin{tabular}{l|ccc}
    \toprule
      Method & Dev Loss & Dev Acc. & Test Acc. \\
    \midrule
      \textbf{original design} & \textbf{0.314} & \textbf{0.906} & \textbf{0.867} \\ 
      - surface form           & 0.318 & 0.900 & 0.848 \\ 
      - KG structure           & 0.331 & 0.896 & 0.847 \\ 
      - description text       & 1.170 & 0.630 & 0.619 \\ 
    \bottomrule
\end{tabular}
\end{table}

The results, as listed in Table \ref{tab:nn-results}, suggest that text feature from the entity description contributes the most to the performance. This coincides with previous findings that simple pattern-matching from the description text could perform reasonably well for typing \cite{kliegr2015linked}. The other two features, surface form and KG structure, also have positive effects.

\subsection{Outlier Detection and Classification}
Outlier detection and classification methods for typing error detection are conducted on a per-type basis, so we evaluated these methods on the fine-grained \textbf{DBpedia-F} dataset and report the macro average precision, recall, F1 of all types. We also report the mean average precision (MAP) to exclude the effect of threshold choice. The results are compiled in Table \ref{tab:fine-results}. We did not report the precision and recall for \emph{Representation Learning + LOF} as the scores are too low with the default threshold, and MAP is sufficient for comparison. We used the method of Caminhas \cite{caminhas2019detecting} as the example for classification method and directly used the results reported in the thesis in our table (marked with $^\star$). This method used the singular value decomposition (SVD) of the one-hot entity-property matrix as property embeddings, and concatenated property embedding and Wikipedia2Vec embedding as the input features. The classifier used was a nearest-centroid classifier, and gold labels were obtained by linking with the LHD dataset \cite{kliegr2015linked}.

\begin{table}[htb]
    \caption{Results of outlier detection and classification methods for typing error detection.}
    \label{tab:fine-results}
    \begin{tabular}{l|cccc}
      \toprule
        Method & Prec. & Recall & F1 & MAP \\
      \midrule
        \textbf{ReprLearning + IF}   & 0.697 & 0.288 & 0.359 & 0.669\\ 
        ReprLearning + LOF  & -     & -     & -     & 0.445\\ 
        Wikipedia2Vec + IF  & 0.593 & 0.170 & 0.222 & 0.585\\ 
        RDF2Vec + IF        & 0.470 & 0.259 & 0.261 & 0.462\\
        Random              & 0.386 & 0.513 & 0.395 & 0.421 \\
        Classification \cite{caminhas2019detecting}  & 0.80$^\star$ & 0.94$^\star$ & 0.86$^\star$ & - \\
      \bottomrule
  \end{tabular}
\end{table}

Overall, the embeddings learned by the representation learning network we proposed worked better than using Wikipedia2Vec or RDF2Vec individually, and Isolation Forest (IF) is more effective than Local Outlier Factor (LOF). However, the performance of outlier detection methods are still poor, with the maximum F1 score lower than the random baseline. As mentioned previously, outlier detection algorithms have assumptions on the density of normal data. However, these assumptions often break in factual KGs as we described in Section \ref{sec:error-rate}, since outliers are sometimes the majority. And we also lack a feasible way to tell which types satisfy the density assumptions, so it is also not easy to apply outlier detection on selected types.

For classification methods, the results appear to be promising. However, they face scalability issues because it is labour-intensive to acquire sufficient amount of labelled data to train good classifiers when the number of types is large, as discussed in Section \ref{sec:classification}. Besides, training large-scale classifiers to ``re-type'' every entity is doing much more than just typing error detection.

\subsection{Discussion}
We have quantitatively compared all four paradigms for typing error detection. All four paradigms have limitations, but we could conclude that the most feasible and effective solution to typing errors is semi-supervised noise models. This is because this paradigm could simultaneously leverage information from entity data points, gold labels, and noisy labels, and hence only requires a minimum amount of human intervention to achieve good performance. In the mean time, outlier detection methods are only applicable if it could be verified that the entity type of interest only contains a very small portion of errors, and classification is only feasible if a good source of supervision is available.

With so many recent research projects relying on DBpedia, how severe is the impact of the typing errors? Although noise are proven to be beneficial for neural networks under certain circumstances \cite{an1996effects}, we believe that this is not the case here. Our experiments have shown a clear (30\%) performance degradation when training our entity typing network with only noisy data. The signal-to-noise ratio (SNR) in factual KGs like DBpedia might be too low and is indeed causing harms. Therefore, it is not appropriate to ignore typing errors when using KGs as datasets, and error detection or correction methods should be considered.

%% file: review.tex
\section{Related Work}
\subsection{KG Assessment}
General reviews on knowledge graphs such as the one by Ringler \cite{ringler2017one} only compared the basic statistics like the size but did not went deep into the quality. Zaveri el al. \cite{zaveri2013user} performed a quantitative assessment of DBpedia, and came up with 4 major issues with it including accuracy. Paulheim and Bizer used the statistical distributions of properties as an indicator of erroneous statements \cite{paulheim2014improving}, and have applied this idea in production for creating the DBpedia 3.9, while our dataset was created from a later version with these improvements. Ma et al. \cite{ma2014learning} came up with the disjoint axiom to estimate the number of typing errors in KGs, stating that an entity should not have two disjoint types. This method was also used by Caminhas \cite{caminhas2019detecting} for error rate estimation. However, this method could only estimate a lower bound of error rate as entities with compatible type labels could also contain errors \cite{caminhas2019detecting}. Users of KGs also paid little attention to the quality of them. For example, Dai et al. \cite{dai2015semi} described DBpedia to have ``no duplication or tainting issues'' (page 7) which is contrary to our findings.

\subsection{Entity Typing and Hypernym Detection}
Entity typing is a task closely related to typing error detection. But generally, entity typing aims at predicting the type of an entity from only a natural language text, while typing error detection could utilize more heterogeneous information like KG structure, noisy label, and other entities with the same type. This is the main difference between our typing network and other neural networks specially designed for entity typing \cite{xu2018neural}. Hypernym detection aims at finding the hypernyms (usually type names) of words, and common methods include training hierarchical embeddings \cite{chang2018distributional} or performing pattern matching \cite{kliegr2015linked}. But our goal is closer to cleansing existing type-word pairs instead of creating new ones.

\subsection{Taxonomy Cleaning}
Recent advances in representation learning has been inspiring research on taxonomy cleaning. For example, similar to our entity typing network, Ren et al. proposed an embedding method that combines information from both text corpus and knowledge graph that claimed that the embedding is resistant to noise \cite{ren2016label}. But as our experiments showed, the use of noise model, VAT and active learning could provide additional help to learning embeddings. Aly et al. \cite{aly2019every} used hyperbolic embeddings to refine taxonomy, but this method is hard to be transferred to our case as it requires high-quality hypernym-hyponym pairs.

\subsection{Learning with Noise}
Goldberger et al. \cite{goldberger2016training} were one of the earliest groups of authors to introduce noise models to the training of neural networks, and this method was applied to NLP by Jindal et al. \cite{jindal2019effective}. Kremer et al. \cite{kremer2018robust} further combines noise models with active learning. However, these work were only evaluated on synthesized data that may not represent label noise in real world like KGs. Our work, on the contrary, is one of the first to apply learning with noise on a real noisy dataset.

%% file: conclusion.tex
\section{Conclusion}
In this study, we exhaustively reviewed all the available paradigms for typing error detection in KGs, and concluded that semi-supervised noise models are the most feasible solution. Under that paradigm, we proposed our method that is optimized to use heterogeneous information from multiple sources. Our opinion is that typing errors in KGs especially DBpedia is a severe problem, and methods such as ours should be deployed when using typing information from DBpedia. Beside errors in entity-type pairs, there are other issues with the DBpedia taxonomy as described in Section \ref{sec:dbpedia-taxonomy-issues}, and we leave the detailed analysis and probable solutions of those issues as future work.